\documentclass[twocolumn,showpacs,preprintnumbers,amsmath,amssymb]{revtex4}

\usepackage{graphicx}
\usepackage{dcolumn}
\usepackage{bm}
\usepackage{psfrag}
\usepackage{epsfig}
\def\dsize{\displaystyle}

\begin{document}


\title{Heat flux modulation in \\domino dynamo model}

\author{Maxim Reshetnyak}%
 \email{m.reshetnyak@gmail.com}
\affiliation{%
Institute of the Physics of the Earth, Russian Academy of Sciences,  Moscow, Russia
}%
\author{Pavel Hejda}
 \affiliation{Institute of Geophysics, Czech Academy of Sciences,  Prague, Czech Republic
}

\date{\today}

\begin{abstract}
Using the domino dynamo model, we show how variations of the heat flux at the core-mantle boundary change the frequency of 
 geomagnetic field reversals. In fact, we are able to demonstrate the effect known from  the modern 3D planetary dynamo models using 
an  ensemble of  interacting spins, which obey equations of the Langevin type with a random force. We also consider 
 applications to the giant planets and offer explanations of some specific episodes of the geomagnetic field in the past.
\end{abstract}

\maketitle

Generation of the planetary magnetic fields is a subject of the dynamo theory, which describes successive
  transformations of thermal and gravitational energy, concerning  compositional convection, to  
 energy of kinetic motions of the conductive liquid and then to the energy of the magnetic 
 field~\cite{bib:RH2004}. Modern dynamo models include partial differential equations of thermal and compositional 
 convection as well as the induction equation for the magnetic field, which some reasons should be three dimensional~\cite{bib:J11}. 
Although, due to the finite 
  conductivity of the Earth's mantle, observations of the geomagnetic field at the Earth's surface 
  are bounded by the first thirteen
  harmonics in the spherical function decomposition, one needs 
 small-scale resolution down to $\sim 10^{-8}\, L$,  to provide the necessary force balance in the core.
Here $L\approx 2.2\, 10^6$m is the scale of the liquid core. This difficulty is caused by the huge
  hydrodynamic Reynolds number $Re\sim 10^9$ as well as by the strong anisotropy of  convection~\cite{bib:HR9} 
 due to the
  geostropic state in the core \cite{bib:Pedl}. Convection in the core is cyclonic. The cyclones and anticyclones 
  are aligned with the axis of rotation, and their scale is much smaller than their length.
   As a result one needs very efficient computer resources to produce regimes 
in the desired asymptotic limit required for 
 geodynamo simulations with grids $128^3$ and more, which is a challenge even for modern supercomputers.

In spite of these technical problems, 3D  dynamo models  successfully mimic various features of the modern and 
 ancient magnetic field including the reversals~\cite{bib:Christ2002}. 
One of the important results of the dynamo theory is that the frequency of the reversals depends
  on the  spatial distribution of the heat flux at the outer boundary of the liquid 
core ~\cite{bib:GR1999}. In particular, 
 the  authors have shown  that the increase of the heat flux along the axis of rotation leads to 
 increase of the axial symmetry of the whole system and stops reversals.
  In a sense, the thermal trap of reversals occurs.
 On the contrary, the decrease of the thermal flux at high latitudes leads to the chaotic behaviour of the magnetic dipole
  accompanied by frequent reversals, which is closely connected with the upsetting of the geostrophic balance and 
 predominance of the radial (in an incompressible medium in which the parameters depend only on the radius)
 Archemedean forces. It looks attractive to obtain this result using toy dynamo models (such as the Rikitake and Lorentz models), 
 which can provide  extensive 
 statistics and obviousness of the results, and 
  simulate  practically 
 instantly, using just home PC, the number of reversals and excursions of the same order as 
 known from paleomagnetism. The random force, which imitates the small-scale unresolved
  fluctuations, was included in some of these models~\cite{bib:Hoyng}. 

For this reason we have  selected  the domino dynamo model~\cite{bib:NMS0, bib:NMS},
 which is an extension of the Ising-Heisenberg XY-models of interacting magnetic spins, for more details of the history 
 of the problem and classification refer to~\cite{bib:Stanley}.

The main idea of the domino model is to consider a system of $N$
 interacting spins 
  ${\bf S}_i,\, i=1\dots N$, in media rotating  with angular velocity  ${\bm\Omega}=(0,\, 1)$ .
 The spins are located over a ring, are of unit length and can vary angle $\theta$
  from the axis of rotation  on  time $t$, so that  ${\bf S}_i=(\sin\theta_i,\, \cos\theta_i)$. 
 Each spin ${\bf S}_i$ is forced by a  random force, effective friction, as well as by  the 
closest  neighbouring spins 
 ${\bf S}_{i-1}$ and ${\bf S}_{i+1}$. 

Following~\cite{bib:NMS0}, we introduce kinetic  $K$ and potential $U$ energies of the system:
\begin{equation}
\begin{array}{l}\dsize
K(t)={1\over 2}\sum\limits_{i=1}^N\dot{\theta}_i^2,     \\ \dsize
 U(t)=\gamma\sum\limits_{i=1}^N \left({\bm \Omega\cdot {\bf S}_i}
\right)^2+ 
\lambda \sum\limits_{n=1}^N 
\left(\bf{\bf S}_{i}\cdot {\bf S}_{i+1}\right).
\end{array}
\label{energy}
\end{equation}
The Lagrangian of the system then takes the form  ${\cal L}=K-U$. 
 Making the transition to the Lagrange equations,  
 adding friction proportional to
  $\dot{\theta}$ and the random force  $\chi$, 
\begin{equation}
{\partial \over \partial t}
{\partial{\cal L} \over \partial \dot{\theta}}
=
{\partial {\cal L} \over \partial \theta}-
\kappa\,\dot{\theta}+{\epsilon \chi\over \sqrt{\tau}},
\label{langevin}
\end{equation}
 leads to the system of Langevin-type equations~\cite{bib:NMS0}:  
\begin{equation}
\begin{array}{l}\dsize
\ddot{\theta}_i-2\gamma\cos\theta_i\sin\theta_i+\lambda
\Big[
\cos\theta_i
\Big(
\sin\theta_{i-1}+\sin\theta_{i+1}
\Big)
-  \\ \dsize
\sin\theta_i
\Big(
\cos\theta_{i-1}+
\cos\theta_{i+1}
\Big)
\Big]  \dsize 
+\kappa\, \dot{\theta}_i+{\epsilon \chi_i\over \sqrt{\tau}}=0,\\
\dsize
  \theta_0=\theta_N,\, \theta_{N+1}=\theta_1,\, i=1\dots N,
\end{array}
\label{final}
\end{equation}
where $\gamma$, $\lambda$, $\kappa$, $\epsilon$, $\tau$ are constants.
 The measure of synchronization of the spins along the axis of rotation 
 \begin{equation}
\begin{array}{l}\dsize
M(t)={1\over N}\sum\limits_{i=1}^N\cos\theta_i(t)  
\end{array}
\label{Mdef}
\end{equation}
will be considered to be the total axial magnetic moment.
{\Huge
\begin{figure}[th]
\vskip -0.5cm
\psfrag{t}{ $t$}
\psfrag{Pol}{$M$}
\psfrag{1a}{\Large a}
\hskip -0.5cm \epsfig{figure=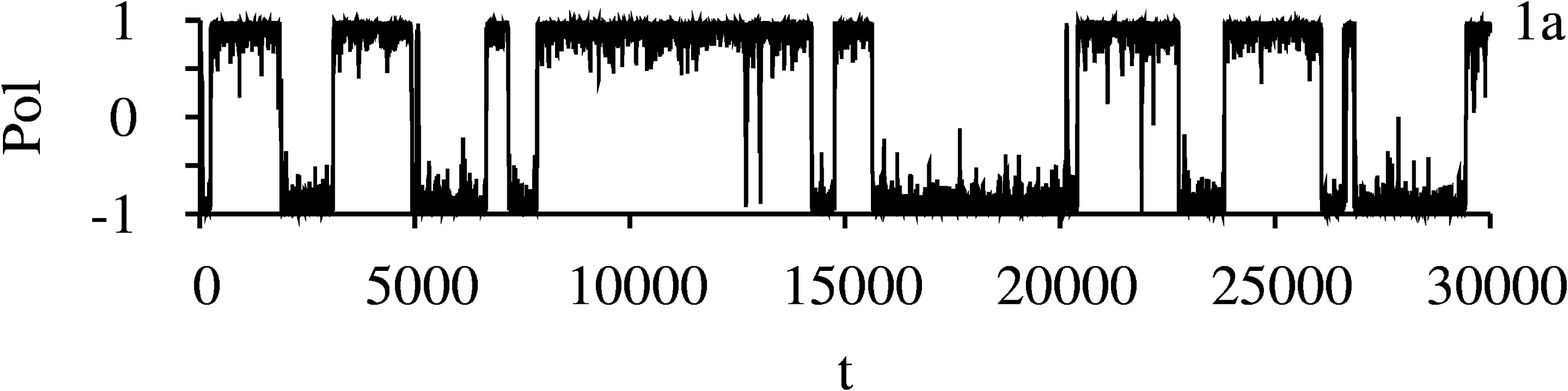,width=9cm}
\vskip 0.1cm
\psfrag{1a}{\Large b}
\hskip -0.5cm \epsfig{figure=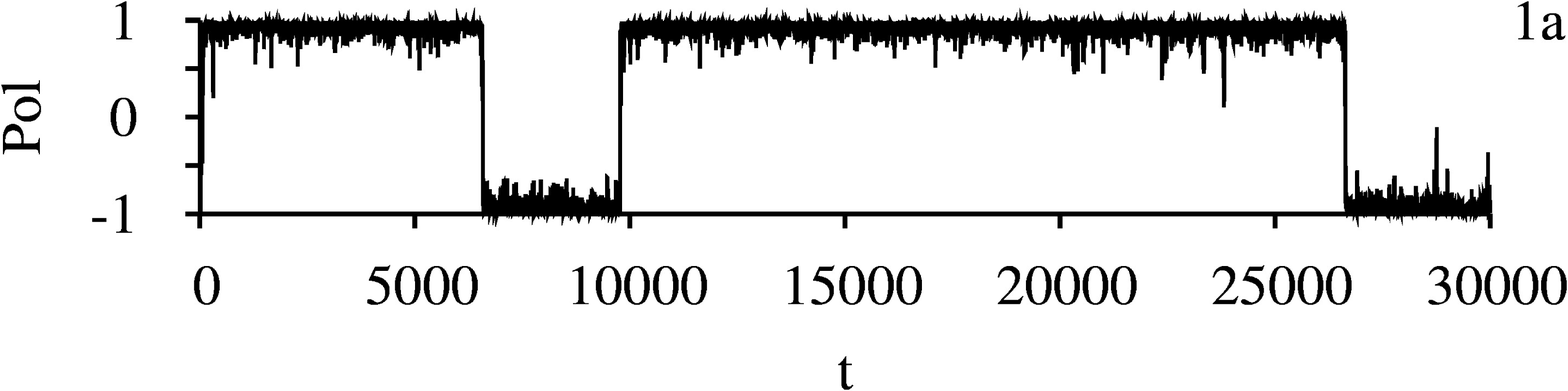,width=9cm}
\vskip 0.1cm
\psfrag{1a}{\Large c}
\hskip -0.5cm \epsfig{figure=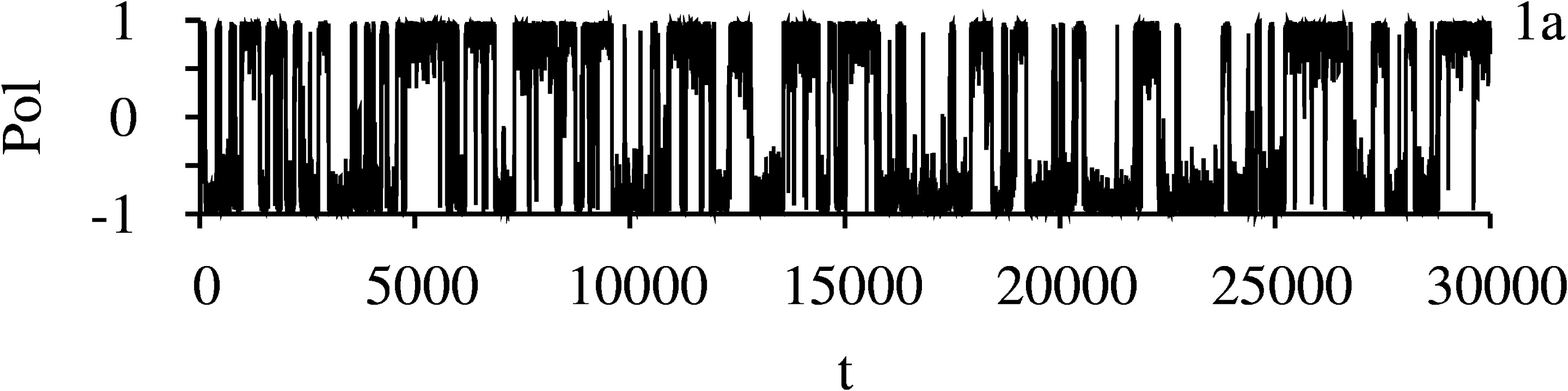,width=9cm}
\vskip 0.1cm
\psfrag{Pol}{$M$, $M_e$}
\psfrag{1a}{\Large d}
\hskip -0.5cm \epsfig{figure=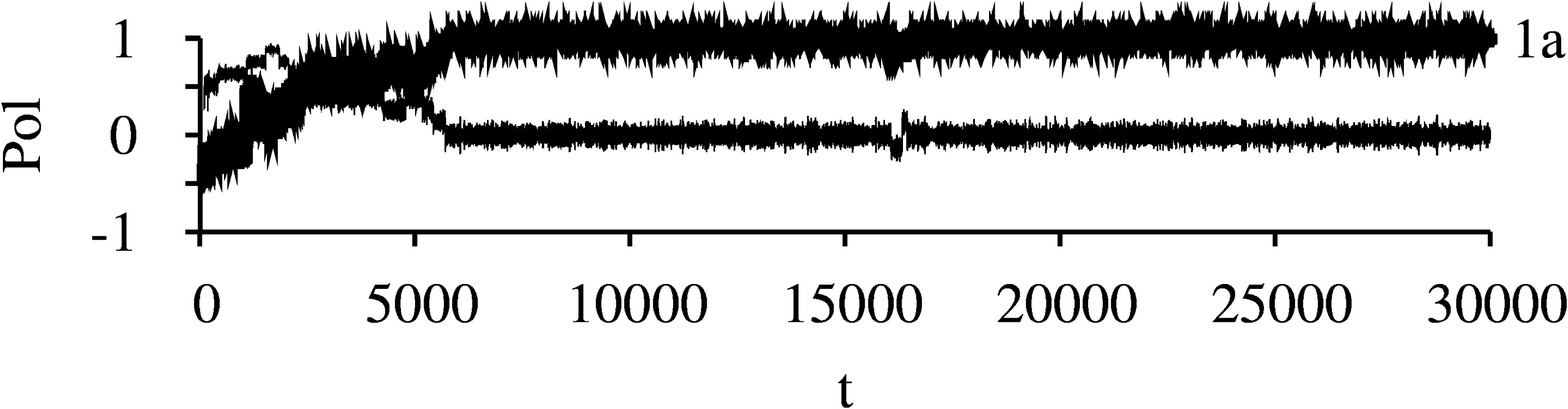,width=9cm}
\vskip 0.1cm
\psfrag{Pol}{$M$}
\psfrag{1a}{\Large e}
\hskip -0.5cm \epsfig{figure=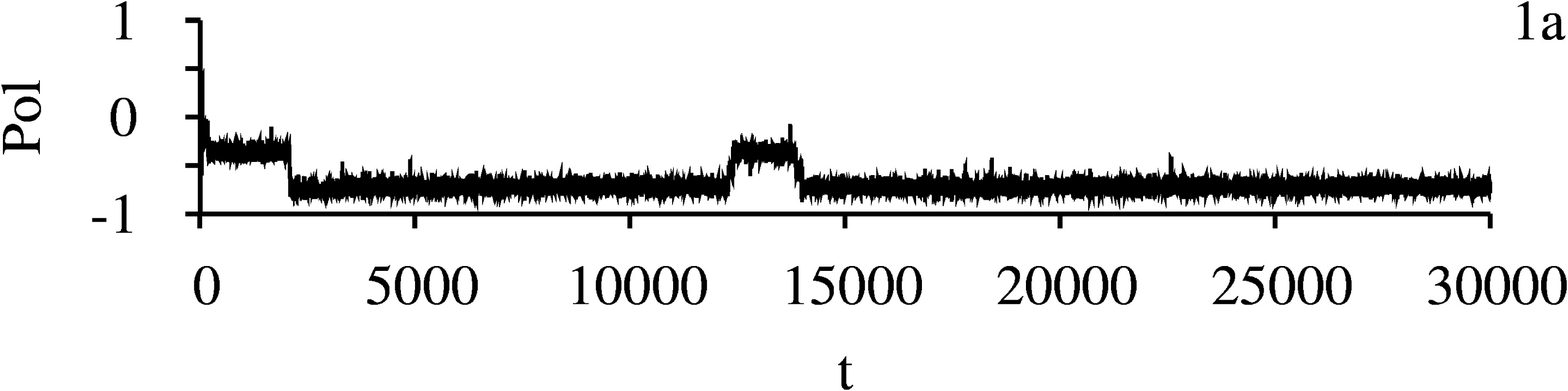,width=9cm}
%
\vskip -0.4cm
 \caption{
Evolution in time of $M$: (a) forthe  purely magnetic system ($C_\psi=0$), (b)  $C_\psi=0.5$,  (c) $C_\psi=-0.5$  for $\psi=-\cos^2\theta$, 
(d) $C_\psi=10$,  (e) $C_\psi=-9$ for  $\psi=-\cos^2 2\theta$. At (d) the thick line corresponds to $M$ and the thin line to 
$M_e$.}
\label{fig1}
\end{figure}
}

Integrating   (\ref{final}) in time for small $N$ 
 one can arrive at quite diverse dynamics of  $M$, 
 very close (for some special choice of parameters) to paleomagnetic observations~\cite{bib:Jacobs}, including the 
   periods of the random and frequent reversals.
It was shown in~\cite{bib:NMS0} that large fluctuation of a single spin can be successively transferred to the neighbouring spins,
which fluctuate until all spins reverse their polarity. This dynamo effect was an inspiration for the name "domino mode".l

In all simulations  we have  used, similarly to~\cite{bib:NMS0}, values of parameters 
 $\gamma=-1$, $\lambda=-2$, $\kappa=0.1$, $\epsilon=0.65$, $\tau=10^{-2}$, $N=8$, 
 with random normal 
  $\chi_i$, with zero mean values and unit dispersions, so that $\chi_i$ was updated at every time step equal to $\tau$.
    As follows from~\cite{bib:NMS0}the  evolution of $M$ depends slightly on the form of the random forcing $\chi$
  and the remaining parameters can be easily selected in such a way as to provide similarity with observations.
{\Huge
\begin{figure}[th]
\vskip -0.0cm
\psfrag{SS1}{$M$}
\psfrag{S1}{$\cos\theta_i$}
\hskip -0.7cm \epsfig{figure=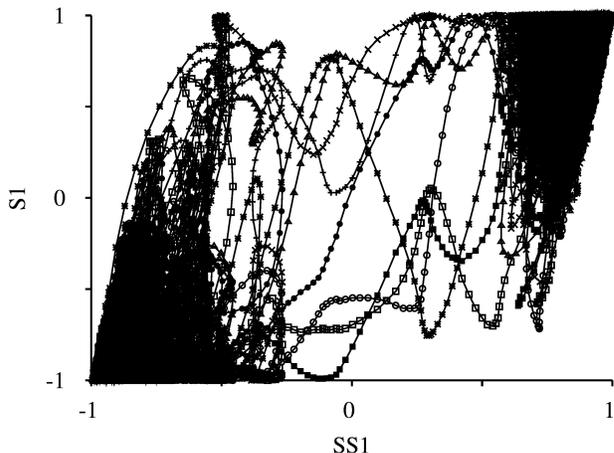,width=9cm}
%
\vskip -0.5cm
 \caption{
Plots of $\cos\theta_i$ with $i=1\dots N$,  versus $M$ for reversal in Fig.1a during the 
 time interval $t=10^3\mbox{--}3\cdot  10^3$.
}
\label{fig2}
\end{figure}
}
The typical behaviour of the axial dipole is presented in  Fig.\ref{fig1}a, where we observe 
   several reversals at irregular time intervals.
  This process is accompanied by a short drop of $M$ to nearly zero and followed by rapid recover,
  which  in geomagnetism is referred to as excursions of the magnetic field.
One can find a thorough analysis of this system    
  in~\cite{bib:NMS0}; 
 here we only emphasize one important in geomagnetism point. Up to now observations do not clearly
indicate,
 if the geomagnetic dipole just rotates during the reversal without decreasing the amplitude, 
orif it decreases  and then recovers with the opposite sign. To check these scenarios we 
present the evolution of the individual
 spins during the reversal, see Fig.\ref{fig2}. There are  quite large deviations of the individual spins
 from the mean value $M$ at the moment of reversal ($M=0$). This means that the decrease of $M$ is caused by the 
 desynchronization of the spins rather than by the coherent rotation of the spins. Note that the typical
  time of the reversal is much larger than the time step. The points on the lines correspond 
 to every $10^{th}$ time step in the simulations.
  The other point is that the minimal time $\sim N\, \tau/2$
  of propagation of the disturbance from spin ${\bf S}_i$ through 
   half the circle  
  is also smaller than the typical reversal time. This scenario is also supported by the 3D 
 simulations, where the spots of the magnetic field with opposite polarities co-exist at the 
 core-mantle boundary  during the reversals.

We now extend the concept of the spin from the purely magnetic system  to the whole cyclone system, 
 including its hydrodynamics, and we introduce correction   $\Psi$  to potential energy 
  $U$ (\ref{energy}), which 
 takes into account the  heterogeneity of the thermal flux. 
 For the purely magnetic problem, it corresponds to the electromagnetic interaction of the cyclone with
  the conducting mantle, the $D"$ layer.  
 The new effective force  $\dsize F_i=-{\partial \Psi_i\over \partial \theta_i}$ ,
 whose influence on the behaviour of  $M(t)$  will be considered in the rest of the paper, then appears in the right-hand side of  (\ref{final}).

Let  $\Psi(t,\, \theta)=C_\psi \, \psi(t,\, \theta)$,  where $C_\psi$ is a constant,
 and the spatial distribution of the potential is given by $\psi=-\cos^2\theta$.  
 Then, 
$C_\psi>0$ 
 corresponds to the stable state in  the polar regions,  
 $\theta=0,\, \pi$, and 
 the appearing force  
$F=-\sin 2\theta$, acting 
  on the cyclones, is directed towards the poles. This regime corresponds to the increase of the thermal flux near the poles
 that causes the stretching of the cyclone along the axis of rotation. 
  In Fig.\ref{fig1} 
  we demonstrate the effective influence of $F$ on  $M$.

The increase of the thermal flux along the axis of rotation leads to the partial  suppression of the reversals
  of the field, Fig.\ref{fig1}b. 
 Note that, for the chosen potential barrier $\psi$, the
  dependence of  $F$  is equal to  the 
 $\gamma$-term 
  in  (\ref{final}): 
the  increase of the thermal flux at the poles leads to the effective increase of rotation and amplification of geostrophy,
 caused by the  rapid daily rotation of the planet.
  Our results are in agreement with the 3D simulations, see Fig.1d in~\cite{bib:GR1999}.  
   The further increase of $C_\psi$ ($C_\psi=2$) leads to the total stop of the reversals. It is more interesting that, 
 using even larger $C_\psi> 10$, one arrive at regimes with a nearly 
 constant in time $|M|\le 1$ defined by the initial 
 distribution of ${\bf S}_i$. In other  words, the super flux at the poles can fix the spins which are still not coherent. 
  There is some evidence~\cite{bib:PV1,bib:PV2} that the geomagnetic dipole in the past couldhave migrated from 
  the usual position near the geographic poles to some stable state in the middle latitudes. Within the framework of our model,
 we can explain this phenomenon by the thermal super flux at the poles. Later we will discuss some other scenarios which yield similar results.

For negative  $C_\psi$,
 when the geostrophy breaks due to the relative intensification of convection in the equatorial plane,
 we get the opposite result, see Fig.\ref{fig1}c: the regime of the frequent reversals observed in
 Fig.1c in~\cite{bib:GR1999}. 
  In this  case force 
 $F$  is directed from the poles and the equilibrium point at  the poles becomes unstable. 
  The new minimum of the potential energy at the equator leads to the appearance of a new attractor, so that 
 for $C_\psi=-5$ the axial dipole fluctuates with  zero mean value and maximal 
   amplitude  $M\sim  0.4$. This state corresponds to the equatorial dipole
\begin{equation}
\begin{array}{l}\dsize
M_e(t)={1\over N}\sum\limits_{i=1}^N\sin\theta_i(t),   
\end{array}
\label{Mdef1}
\end{equation}
 so that $|M_e|\sim 1$ and does not undergo reversal. Similar behaviour of the magnetic dipole is observed on 
  Neptune and Uranus;  for more details see,  e.g.~\cite{bib:CHR}.

Now we consider  $\psi=-\cos^2 2 \theta$ for which the corresponding 
   force $F=-2\sin 4 \theta$ changes sign
  in each hemisphere, see the example with the second-order zonal spherical harmonic  
   $\sim P_2^0$, where  $P_l^m$ is associated Legendre polynomials, in Fig.1e in~\cite{bib:GR1999}.
 
 For $C_\psi>0$, 
 the potential barrier is  in the middle latitudes, which prohibits the reversals as observed in  
 Fig.1e in~\cite{bib:GR1999}. In one of our runs only one reversal was observed at $C_\psi=2$. 
 The other important point is the existence of the stable point at the equator, where $M=0$. We could then assume a 
 regime with two  attractors: near the poles and at the equator. This regime is really observed in 
 Fig.\ref{fig1}d. The inverse transition was not observed. Inspection of the state with small $M$ at the beginning
 of the run  leads to 
 a  quite large estimate of the equatorial dipole amplitude $M_e$, see~Fig.\ref{fig1}d. Thus, in principle, this regime can also  be 
  related to the giant planets' dynamo as.

The last example corresponds to the case when $C_\psi<0$. 
 Here in addition to the attractors at the poles (related to rotation  $\Omega$), 
 two new attractors appear at middle latitudes. The variation of $C_\psi$, 
 leads to regimes $C_\psi=-1$ with frequent reversals, observed in Fig.1f 
 in~\cite{bib:GR1999}. Moreover, we can get regimes, see Fig.\ref{fig1}e,
   when the magnetic pole  stays at high latitudes, however, 
 $|M|\ne 1$ (the partial synchronization of the spins). There are some jumps to the unstable state $M=0$
 with the dipole  at the equator. Note that the  decrease of the spatial scale of $\psi$ leads to the increase of 
 its amplitude $C_\psi$.

Obviously, as the interaction between the cyclones is  certainly  more complex~\cite{bib:Kag} than 
in (\ref{energy}) and the mechanism of magnetic field generation in cyclonic  turbulence
 is outside the scope of the domino model,  our results should be used with caution. 
 However, it is remarkable that even such a simple model 
  provides quite a wide range of effects, known from very sophisticated models and observations. 
This, of course, does not
 exclude other explanations and scenarios of the considered phenomena.

\end{document}